\documentclass[english,aps,prl,twocolumn,groupaddress,showpacs,reprint,amsmath,amssymb]{revtex4}
\usepackage{bm}
\usepackage{amsmath}
\usepackage{amssymb}
\usepackage{graphicx}
\usepackage{txfonts}

\usepackage[colorlinks=true,urlcolor=blue,citecolor=blue,linkcolor=blue,breaklinks=true]{hyperref}
\begin{document}

\title{Interaction-Driven Topological Insulator in Fermionic Cold Atoms on an Optical Lattice:\\
A Design with a Density Functional Formalism}

\author{Sota Kitamura}
\affiliation{Department of Physics, University of Tokyo, Hongo, Tokyo 113-0033,
Japan}

\author{Naoto Tsuji}
\affiliation{Department of Physics, University of Tokyo, Hongo, Tokyo 113-0033,
Japan}

\author{Hideo Aoki}
\affiliation{Department of Physics, University of Tokyo, Hongo, Tokyo 113-0033,
Japan}

\date{\today}
\begin{abstract}
We design an interaction-driven topological insulator for fermionic
cold atoms in an optical lattice, that is, we pose the question of whether
we can realize in a continuous space a spontaneous symmetry breaking
induced by the inter-atom interaction into a topological Chern insulator.
Such a state, sometimes called a ``topological Mott 
insulator", has
yet to be realized in solid-state systems, since this requires, in
the tight-binding model, large offsite interactions on top of a small
onsite interaction. Here we overcome the difficulty by introducing
a spin-dependent potential, where a spin-selective occupation of fermions
in $A$ and $B$ sublattices makes the onsite interaction 
Pauli-forbidden, 
while a sizeable inter-site interaction is achieved by a shallow optical
potential with a large overlap between neighboring Wannier orbitals.
This puts the system away from the tight-binding model, so that we
adopt the density functional theory for cold-atoms, here extended
to accommodate non-collinear spin structures emerging in the
topological regime, to quantitatively demonstrate the phase transition
to the topological Mott insulator.
\end{abstract}

\pacs{67.85.-d, 73.43.Nq, 71.30.+h, 71.15.Mb}

\maketitle

\textit{Introduction.---} There is a growing fascinations with topological
phases in condensed-matter physics~\cite{Qi2011,Hasan2010}. The topological
systems are characterized by various topological invariants~\cite{Schnyder2008,Kitaev2009},
e.g., Chern numbers~\cite{Thouless1982,Haldane1988}, as exemplified
by the quantum Hall system, and $Z_{2}$ numbers~\cite{Kane2005,Konig2007}
for the topological insulator.  Such topologically nontrivial phases
emerge from one-body physics: a nonzero Chern number arises when
the time-reversal symmetry (TRS) is broken, e.g., by a strong external magnetic
field. A nonzero $Z_{2}$ number can be realized without breaking
the TRS, while typically a large spin-orbit coupling
is required.

Recently, a class of topological phases was proposed 
where the one-body terms required to make systems topological
 are absent 
but envisaged to emerge from a spontaneous symmetry breaking (SSB) due to many-body interactions~\cite{Raghu2008,Sun2009,Hohenadler2013}.
Such systems, called the ``topological Mott insulator'' (TMI),
accompany interaction-driven loop currents, which act as an effective
magnetic field or spin-orbit coupling. The TMI has been suggested to
arise from repulsive inter-site interactions, mainly from Fock-term
contributions, and the existence of such anomalous topological phases
has been suggested for the 
extended Hubbard model on
various tight-binding lattice models~\cite{Raghu2008,Sun2009,Zhang2009,Wen2010,Kurita2011}.
However, despite several proposals~\cite{Zhang2009,Kurita2011,Sun2012,Dauphin2012,Manzardo2014,Dora2014,Liu2014}, condensed-matter realization of such phases has yet to
be achieved.
A crucial difficulty is that the TMI often requires large
inter-site interactions to trigger the desired SSB,
while onsite and other interactions must be suppressed 
to avoid competing instabilities.

On the other hand, 
ultracold atom systems~\cite{Bloch2008,Giorgini2008} in optical lattices
provide a clean and tunable platform for exploring exotic
topological phenomena~\cite{Aidelsburger2011,Duca2014,Jotzu2014}.
In cold-atom systems, where interactions 
controlled by the $s$-wave scattering length can be 
varied with the Feshbach resonance~\cite{Chin2010}, 
the inter-site interactions are too small to realize the 
TMI. Several 
studies~\cite{Raghu2008,Zhang2009,Sun2012,Dauphin2012} propose to
circumvent this difficulty~\cite{Sowinski2013}
by employing, e.g., interactions of molecules or Rydberg atoms. 
However, while these 
setups may indeed induce significant inter-site interactions, 
whether the TMI phase is actually favored over competing and adverse effects will have to be studied.

In the present work we design an optical lattice
system for the interaction-driven topological phase transition, 
where atoms still experience $s$-wave scatterings 
but the transition can emerge.  
A key ingredient is a spin-dependent optical lattice potential,
whose minima yield a spin-selective occupation of fermions in the $A$ and $B$ sublattices 
of a square lattice. This washes out the onsite interaction 
due to Pauli exclusion to make
the nearest-neighbor interaction the leading one~\cite{Ruostekoski2009,Jaksch1998}. 
While this setup, in the tight-binding limit, corresponds to a 
checkerboard 
model studied by
Sun \textit{et al.}~\cite{Sun2009}, the present model in a continuous space requires a wide enough 
breadth of Wannier orbitals for strong inter-site
interactions, which invalidates the tight-binding picture. Thus, we
have definitely to depart from the tight-binding picture, 
so that we employ density functional theory (DFT) for cold-atom systems~\cite{Ma2012}, 
here extended to accommodate non-collinear spin-density functionals
to describe topological spin textures. We then demonstrate quantitatively
that the proposed cold-atom system in a continuous space does indeed
exhibit a topological phase transition from a semimetal to
a Chern insulator with a significant topological gap, as the repulsive interaction is increased.

\textit{Basic idea.---} We consider ultracold
fermions of spin-1/2 in a continuous space in the presence of an
optical lattice potential, with a Hamiltonian
\begin{align}
\hat{H}_{\text{OL}}= & \sum_{\sigma,\sigma^{\prime}}{\int}d\bm{r}\hat{\psi}_{\sigma}^{\dagger}(\bm{r})\Bigl[-\dfrac{\hbar^{2}}{2M}\delta_{\sigma\sigma^{\prime}}\bm{\nabla}^{2}+V_{\sigma\sigma^{\prime}}(\bm{r})\Bigr]\hat{\psi}_{\sigma^{\prime}}(\bm{r})\nonumber \\
 & +{\int}d\bm{r}d\bm{r}^{\prime}\hat{\psi}_{\uparrow}^{\dagger}(\bm{r})\hat{\psi}_{\downarrow}^{\dagger}(\bm{r}^{\prime})U(\bm{r}-\bm{r}^{\prime})\hat{\psi}_{\downarrow}(\bm{r}^{\prime})\hat{\psi}_{\uparrow}(\bm{r}).\label{eq:continuousmodel}
\end{align}
 Here, $\hat{\psi}_{\sigma}(\bm{r})$ is the fermion field
operator with mass $M$, while $V_{\sigma\sigma^{\prime}}(\bm{r})=W(\bm{r})\delta_{\sigma\sigma^{\prime}}+\bm{B}(\bm{r})\cdot\bm{s}_{\sigma\sigma^{\prime}}$
is a spin-dependent optical lattice potential, consisting of a periodic
potential $W(\bm{r})$ and a periodic Zeeman field $\bm{B}(\bm{r})$,
with the Pauli matrix $\bm{s}_{\sigma\sigma^{\prime}}$.  
$U(\bm{r}-\bm{r}^{\prime})$ is 
the hard-core fermion-fermion interaction whose radius coincides with $a_s$, the $s$-wave scattering length.

\begin{figure}
\begin{centering}
\includegraphics[width=1\columnwidth]{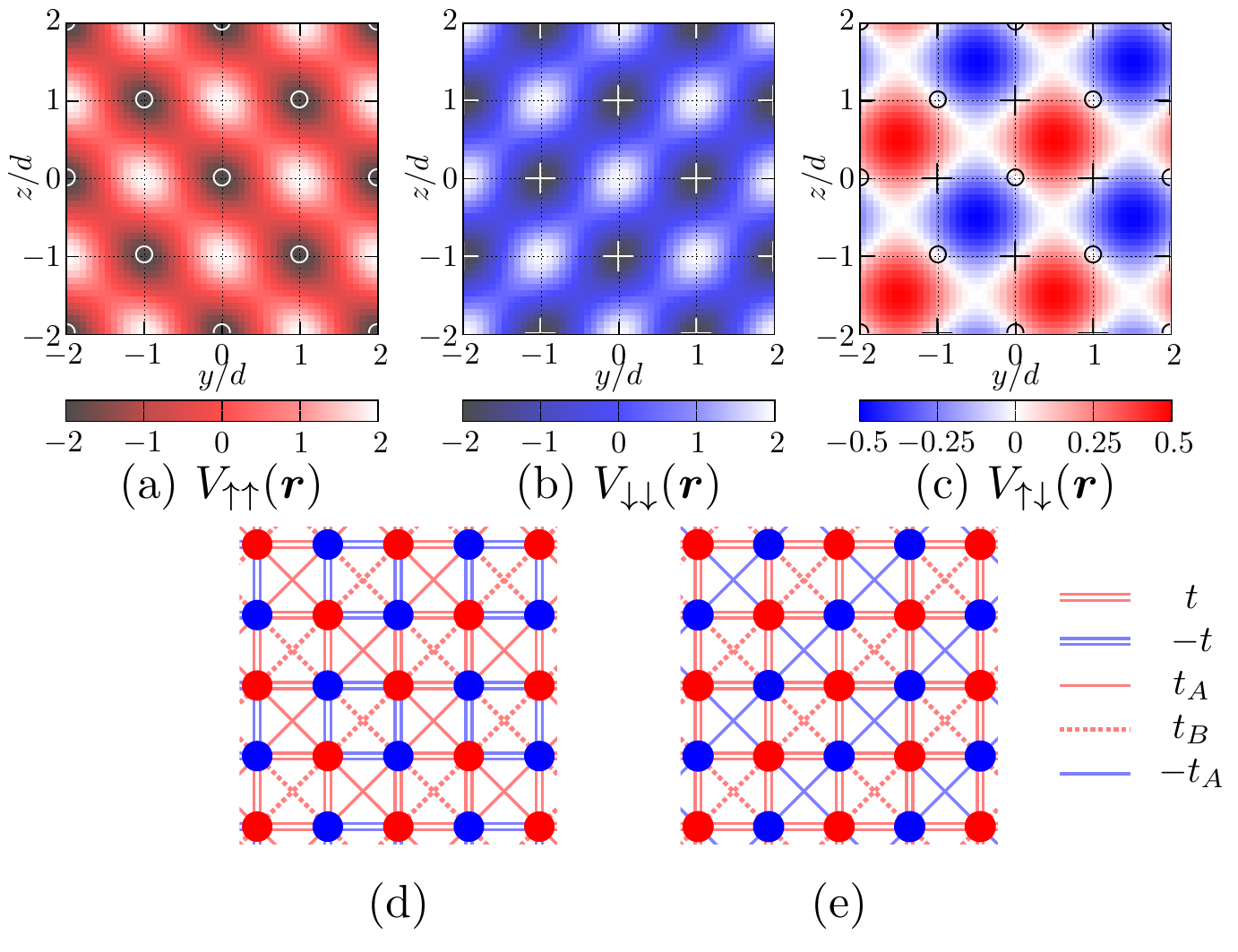}
\par\end{centering}

\caption{\label{fig:hopping} (color online). (a-c) Spatial patterns of the
optical lattice potentials $V_{\uparrow\uparrow}(\bm{r})$ (a), $V_{\downarrow\downarrow}(\bm{r})$ (b),
and $V_{\uparrow\downarrow}(\bm{r})$ (c), given, respectively, in Eqs.~(\ref{eq:uppotential},~\ref{eq:downpotential},~\ref{eq:crosspotential}),
here shown on the $x=0$ plane for $(V_{A},V_{B})=(0.8,1.2)$ in
units of $E_{R}=\hbar^{2}\pi^{2}/(4Md^{2})$. $\circ(+)$ indicates
positions of minima in $V_{\uparrow\uparrow}(V_{\downarrow\downarrow})$.
(d) Spatial pattern of the hopping in the tight-binding limit. Red
(blue) circles represent $A$ ($B$) sites, while red (blue) lines
represent the hopping with positive (negative) amplitudes. (e) A transformed
model.}
\end{figure}

In tight-binding models~\cite{Raghu2008,Sun2009,Zhang2009,Wen2010,Kurita2011},
realization of the TMI is shown to require repulsive 
offsite interactions on a lattice with 
a semimetallic one-body band structure. 
Even in cold atoms with a short-range
interaction, we can generate offsite interactions by employing
a spin-dependent optical potential. For this purpose we adopt
a form of the potential
\begin{gather}
V_{\uparrow\uparrow}(\bm{r})=-V_{A}\cos\dfrac{\pi(y-z)}{d}-V_{B}\cos\dfrac{\pi(y+z)}{d}+V_{T}(x),\label{eq:uppotential}\\
V_{\downarrow\downarrow}(\bm{r})=V_{B}\cos\dfrac{\pi(y-z)}{d}+V_{A}\cos\dfrac{\pi(y+z)}{d}+V_{T}(x),\label{eq:downpotential}\\
V_{\uparrow\downarrow}(\bm{r})=V_{\downarrow\uparrow}(\bm{r})=V_{C}\Bigl(\sin\dfrac{{\pi}y}{d}+\sin\dfrac{{\pi}z}{d}\Bigr),
\label{eq:crosspotential}
\end{gather}
which accomplishes 
sizeable nearest-neighbor (NN) interactions and semimetallic band structures as we indicate below.  Here, $V_{T}(x)$
is a trapping potential along $x$, taken to have a form $V_{T}(x)=V_{x}\sin^{2}({\pi}x/2d)$ with $d$ the lattice constant, 
where a cutoff in the third direction ($x$) to $[-d,d]$ is imposed.  
Spatial patterns at $x=0$ are depicted in Figs.~\ref{fig:hopping}~(a-c).

Let us explain what strategy led us to adopt the form of the potential Eqs.~(\ref{eq:uppotential}-\ref{eq:crosspotential}) (for detailed implementations see the Supplemental Material).  
Although we work in a continuous space for cold atoms, we can first have an intuitive look at how the system would look
in the tight-binding limit. We start with the spin-diagonal
components [Eqs.~(\ref{eq:uppotential},~\ref{eq:downpotential})],
whose positions of minima are made to be spin-dependent unlike ordinary lattice models (see Figs.~\ref{fig:hopping}~(a,~b)): 
up-spin fermions occupy the $A$ sublattice sites of the square lattice while down-spins occupy the $B$ sites.  In other words, we 
can regard the system as that of spinless fermions if we translate
the spin into the sublattice index. 
In this situation onsite interactions simply vanish due to 
the Pauli exclusion principle, so that the leading
interaction is the 
nearest-neighbor $V_1$, which arises from overlapping
tails of neighboring Wannier orbitals.  We note that, while this idea of encoding 
NN interactions is adopted from Ref.~\onlinecite{Ruostekoski2009},
where a kagom\'{e} lattice with NN interactions is described in a
tripod scheme of resonant transitions between atoms 
in three levels corresponding to the three sublattices in the kagom\'{e} lattice,
the present scheme is fairly different as it employs off-resonant
lasers~\cite{Suppl}, although a spin-dependent potential finally 
emerges~\cite{Deutsch1998,Dudarev2004}.

As far as the spin-diagonal lattice potential is concerned, NN hopping is absent despite the large overlap, 
since the system is spin-diagonal and conserves spin. We can then 
induce the hopping by adding the spin-offdiagonal part, Eq.~(\ref{eq:crosspotential}),
as a perturbation, 
which induces
spin flips.
Since Eq.~(\ref{eq:crosspotential}) is a 
staggered  Zeeman field along $x$, 
the hopping amplitude takes a real value with
alternating signs (see Fig.~\ref{fig:hopping}~(c)).
Then, the corresponding tight-binding model with NN interactions 
becomes as depicted in Fig.~\ref{fig:hopping}~(d), where $t_{A}$
($t_{B}$) denotes the hopping through the potential
barriers in Eqs.~(\ref{eq:uppotential},~\ref{eq:downpotential}), while
$t$ denotes those through the off-diagonal part, Eq.~(\ref{eq:crosspotential}).
We can perform a unitary transformation~\cite{Suppl}
to put the model into a simpler one as depicted in Fig.~\ref{fig:hopping}~(e).
The transformed model is 
the so-called checkerboard lattice for spinless fermions
with alternating signs
for the second-neighbor hoppings. The tight-binding model is semimetallic
at half-filling in the non-interacting case, and is theoretically reported
to have a nonzero Chern number even for infinitesimal $V_1$ due
to a spontaneous breaking of the TRS~\cite{Sun2009}.
Hence, we can expect that the present model in the original continuous
space $\hat{H}_{\text{OL}}$ should be promising for realizing the interaction-driven Chern insulator.

In passing, we remark on the symmetry of the present system: the TRS,
defined as a physical symmetry (which inverts spin
directions), is explicitly broken by the Zeeman fields ($B^{z}$
and $B^{x}$). However, the $B^{y}$-component is absent so that the Hamiltonian
is real, and the system has a symmetry against the complex conjugation.
The complex conjugation corresponds to the TRS
in the spinless system (which does not invert sublattice indices),
where the breaking of that symmetry signifies the topological phase
transition\footnote{Namely, the present many-body Hamiltonian has an anti-unitary symmetry
$K$ with $K^{2}=1$ 
(class AI), where topological phases
are absent in two-dimensional non-interacting systems~\cite{Schnyder2008}.
Nevertheless, many-body effects allow the system to be a Chern insulator,
because the Kohn-Sham Hamiltonian of the system
after the SSB
can be class A.}. 
Hence, the ordered phase of the present system should accompany a
(staggered) magnetization along the $y$-axis inducing a magnetic field
along the $y$-axis, i.e., the imaginary part of 
the spin-offdiagonal mean-field potential~\cite{Suppl}. In terms of the spinless tight-binding
model, that translates into a complex NN hopping amplitude, and is
consistent with spontaneous loop currents in the topological phase~\cite{Sun2009}.
In short, the present idea is summarized as follows; (i) by employing
the spin-dependent potential minima, 
a NN interaction arises in a spinless square lattice system 
and (ii) by the staggered Zeeman field
along $x$, a checkerboard pattern of the second-neighbor hopping
with alternating signs is realized, 
which realizes a semimetal in the non-interacting case.

\textit{Method.---} Now, we go back to the original problem in
the continuous space. The whole point is that, although the 
tight-binding limit has a desired form, there is no guarantee that
this reduction is valid. In fact, the situation does call for a careful treatment: 
while we have to employ a shallow lattice
potential for a large overlap between
neighboring Wannier orbitals to enhance the NN interaction, the shallow potential will 
also enhance the longer-distance hoppings and the effects of excited bands,
which may well degrade the desired situation. This is precisely
why we have to employ DFT to directly solve $\hat{H}_{\text{OL}}$.
The exchange-correlation functional for ultracold fermions within
the local spin-density approximation (LSDA) has been formulated in
a pioneering work by Ma \textit{et al.}~\cite{Ma2012}, where it is
reported that the LSDA quantitatively reproduces the total energy of the
shallow optical lattice system as estimated from the diffusion Monte
Carlo method.

If we apply DFT to the present system, however, the potential
$V_{\sigma\sigma^{\prime}}(\bm{r})$ is not spin-diagonal, so that
we have to deal with \textit{non-collinear} spin DFT~\cite{Sandratskii1986,Kubler1988}, 
where the spatial pattern of spin directions is allowed to have general
structures. We adopt here the local approximation.  Then, we can obtain the energy functional from the collinear one given in Ref.~\onlinecite{Ma2012}, $E^{\text{(c)}}_{\text{HXC}}[n_{\uparrow}(\bm{r}),n_{\downarrow}(\bm{r})]$,
by replacing the axis of the collinear spin density $n_{\uparrow,\downarrow}(\bm{r})$ by a spatially-varying one, as is done in electron systems~\cite{Kubler1988,Suppl}.

We can mention that the non-collinear LSDA formalism 
is appropriate to the TMI in cold-atom systems: while for the long-ranged
Coulomb interaction, the Fock term, which is essential for the topological
transition, is nonlocal and 
difficult to capture with the local-density approximation,
the Hartree-Fock energy for the contact interaction can be expressed explicitly within the
non-collinear LSDA functional~\cite{Suppl}.
Further, the topological phase suggested in the checkerboard lattice
is expected to occur in a weakly-correlated regime ($V_1\lesssim|t|$), where DFT
should be adequate.

\textit{Results.---} Now, we present the density-functional results.
We take the periodic boundary condition to employ Bloch wave functions.
The number of $k$-points is taken as $1\times32\times32$, and each
Bloch function is represented by $9\times21\times21$ plane waves.
We set the parameters of the optical lattice potential as $V_{A}=0.8,V_{B}=1.2,V_{C}=0.25$, and $V_{x}=10$, all in units of $E_{R}=\hbar^{2}\pi^{2}/(4Md^{2})$.

\begin{figure}
\begin{centering}
\includegraphics[width=1\columnwidth]{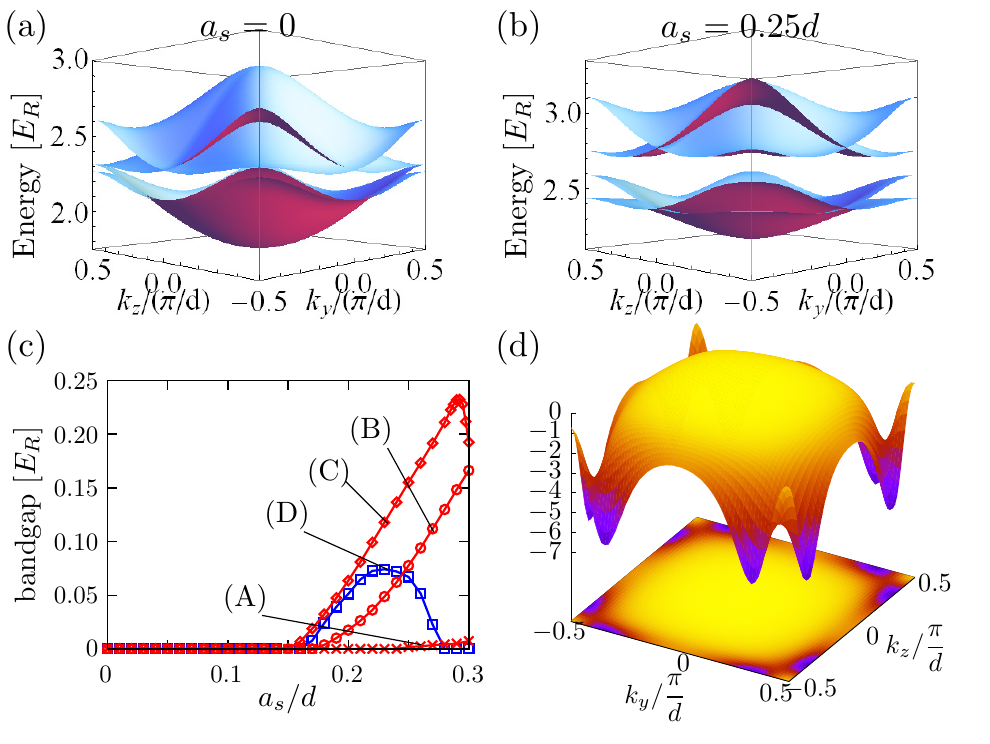}
\par\end{centering}

\caption{\label{fig:bandgap} (color online). Structure of the lowest four
bands, in the non-interacting case (a) and for an interacting system (b)
with $a_{s}=0.25d$. (c) The topological gap of the system against
$a_{s}$. The parameters, all in units of $E_{R}$, are $V_{A}-V_{B}=0.4$ (A),
$-0.4$ (B), and $-0.8$ (C) with $(V_{A}+V_{B},V_{C})=(2,0.25)$ for all
the cases. A round-off of the gap for stronger $a_{s}$ for (C) signifies
an emergence of a site-nematic order (see text). (D) displays the
case of (C) with a magnetic field applied along $x$ with $V_{M}=0.2$.
(d) The Chern density for the lowest two bands in the system is depicted
in (b).}
\end{figure}

\begin{figure}
\begin{centering}
\includegraphics[width=1\columnwidth]{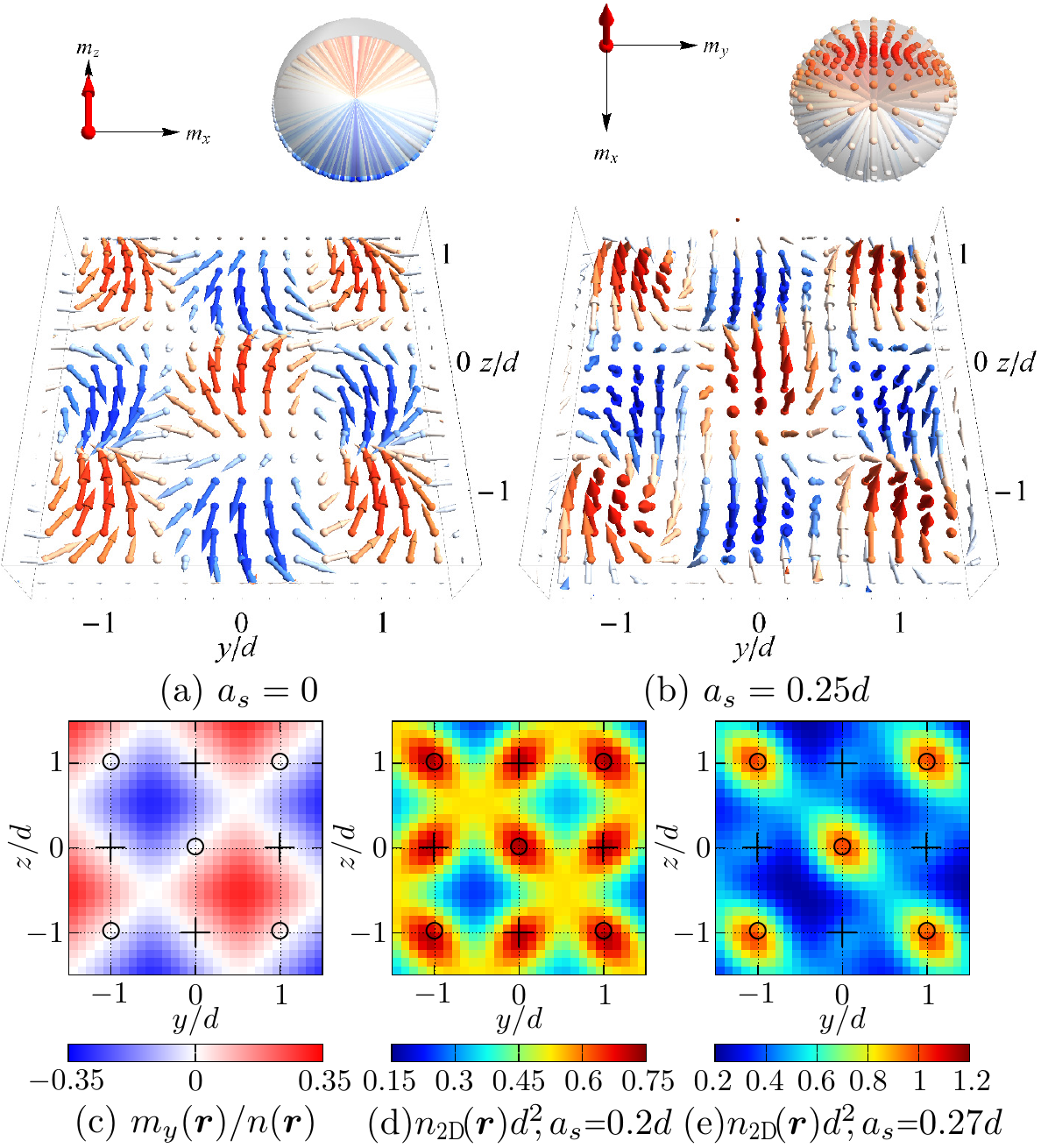}
\par\end{centering}

\caption{\label{fig:topological} (color online) (a, b) Spatial texture of
the magnetization $\bm{m}$ (arrows) for the non-interacting case (a) and for
an interacting system with $a_{s}=0.25d$ (b) with the color representing the
$z$-component magnetization up (red) or down (blue). 
The axes of $\bm{m}$ are taken as indicated 
for clearer 
viewing, and the loci of spins as we sweep the unit
cell are depicted on Bloch spheres in the top insets. 
(c) The $y$-component of the
magnetization (order parameter for the topological phase) for the state
depicted in (b). (d) Spatial pattern of the atomic area density for
the case of Fig.~\ref{fig:bandgap}~(c)~(C) with $a_{s}=0.2d$. (e) The same
as (d) for a larger $a_{s}=0.27d$, for which a site-nematic order
coexists with a topological gap.}
\end{figure}

In the non-interacting case, $a_{s}=0$, we have a band structure
as depicted in Fig.~\ref{fig:bandgap}~(a). Because of the staggered fields,
a unit cell contains four lattice sites,
so that the bottom four bands correspond to those in the tight-binding
model. Two quadratically dispersive bands 
touch 
each other at the corner of the Brillouin zone, which is called a
quadratic band-crossing point (QBCP), associated with the symmetry
against complex conjugation. At half-filling, the lowest two bands
are fully-occupied, and the system is a semimetal. When the interaction
is switched on in Fig.~\ref{fig:bandgap}~(b), however, we can see that
a gap opens at the QBCP, where a SSB makes the
system an insulator. The size of the gap as a function of $a_{s}$
in Fig.~\ref{fig:bandgap}~(c)~(B) shows that the gap grows with the interaction
with a threshold behavior.

The gap is indeed a topological gap, which is verified from the Chern
number~\cite{Fukui2005}. Figure~\ref{fig:bandgap}~(d) shows the Chern
density, summed over the lowest two bands,
where we can see conspicuous magnitudes around the QBCPs. The Chern
number as an integrated value turns out to be $-1$, and we can conclude
that the system is a Chern insulator driven by SSB from a semimetallic phase.

The spin structure of the system in continuous space is depicted in
Figs.~\ref{fig:topological}~(a-c). As we have noted above in the discussion
of the symmetry, the order parameter for the present system is the
staggered magnetization along the $y$-axis. Figure~\ref{fig:topological}~(a)
shows the non-interacting case, where the spatial spin structure comes
from the Zeeman field in the $x\text{-}z$ plane, so that the $y$-component
is trivially absent. In the interacting case, by sharp contrast,
the $y$-component magnetization spontaneously emerges, as is most clearly
seen in the Bloch sphere inset, and we can identify the insulator
to be topological. Accordingly, the spatial behavior of the spins
in the periodic system changes from two-dimensional vortices to three-dimensional
ones (Figs.~\ref{fig:topological}~(b,~c)). Hence, we conclude that the
designed system does indeed realize, in the continuous space, the
mechanism for the emergence of the topological phase conceived for
tight-binding models.

Figure~\ref{fig:topological}~(c) indicates that the order parameter
has large amplitudes around $(y,z)=({\mp}d/2,{\pm}d/2)$, where the
atomic density gives the upper limit for $m_{y}$. Hence, we can enlarge
the topological gap by enhancing the density around these positions.
This can be achieved by controlling the anisotropy of the potential
barrier separating adjacent $A$ sites (or $B$ sites) (i.e., reducing $V_{A}-V_{B}$),
as shown in Figs.~\ref{fig:bandgap}~(c)~(A-C), where $V_{A}-V_{B}$ has
an effect of increasing the gap. Figure~\ref{fig:topological}~(d)
shows that the density around $(y,z)=({\mp}d/2,{\pm}d/2)$ is sizeable~\cite{Suppl}.
We can notice a round-off of the gap at stronger interactions in Fig.~\ref{fig:bandgap}~(c)~(C),
which we identify as coming from another phase transition into a site-nematic
order (i.e., a spontaneous imbalance of the filling between the two
sublattices), which is reported for a tight-binding model in Ref.~\onlinecite{Sun2009}.
The coexistence of the topological and nematic orders occurs above $a_{s}\sim0.23d$
in the present setup (see Fig.~\ref{fig:topological}~(e)).

\textit{Discussion.---} We have neglected some
factors that may work against the topological transition: (i) thermal
fluctuations, (ii) a Zeeman splitting accompanying the Feshbach resonance,
(iii) three-body scattering processes, which can induce an instability
toward a dimerized phase~\cite{Sanner2012,Zintchenko2013}, and (iv)
nonlocal or dynamical correlations dropped in the LSDA. Let us discuss (i) and (ii) in detail.
The critical temperature should have an order of magnitude of the
topological gap, which is scaled by the bandwidth. The bandwidth is
larger in shallower optical lattices, so that we can expect that
the critical temperature can be made accessible. If we introduce a
uniform Zeeman field, $V_{M}s_{\sigma\sigma^{\prime}}^{x}/2$, we
can estimate the upper bound of the magnetic field to be $V_{M}\sim2V_C$,
at which the Zeeman splitting makes $V_{\uparrow\downarrow}(\bm{r})$
non-staggered. As shown in Fig.~\ref{fig:bandgap}~(c)~(D), the topological
order, although reduced, persists in a magnetic field of $V_{M}=0.8V_C$.

Experimental detections 
of topological quantities are crucial for discriminating from 
conventional phase transitions.  While this is challenging, 
various experimental methods are now being 
proposed in general~\cite{Price2012,Dauphin2013,Goldman2013,Alba2013,Deng2014,Lisle2014,Hauke2014,Stamper-Kurn1999,Liu2010,Goldman2012}, which may be applicable to the present system as we discuss in Supplemental Material~\cite{Suppl}.  
The most promising is to observe gapless 
excitations originating from the chiral edge modes with light Bragg spectroscopy as highlighted in Fig.~S3~\cite{Suppl}. 
The method directly measures the dynamical structure factor, which reflects the emergence of the in-gap 
edge modes inherent in the topological 
transition, that can be further endorsed by an edge-selective irradiation.

To summarize, the present design is the first example of a realistic
model in a continuous space that exhibits an interaction-induced SSB toward the Chern insulator. Compared to the other
proposals for the TMI in cold-atom systems~\cite{Raghu2008,Zhang2009,Sun2012},
our proposal has some advantages: it employs only a simple and established
scheme for cold atoms, i.e., the $s$-wave Feshbach resonance and
the electric dipole transition between hyperfine states induced by
off-resonant lasers, along with shallow lattice potentials, which
tend to enhance the transition temperature.

\begin{acknowledgments}
The authors would like to thank P. N. Ma, M. Troyer, and T. Oka for
illuminating discussions. This work is supported by a Grant-in-Aid
for Scientific Research (Grant No.~26247064) from MEXT, and (S.~K.) by
the Advanced Leading Graduate Course for Photon Science (ALPS).
\end{acknowledgments}

\end{document}


\title{Supplemental Material for ``An interaction-driven topological
insulator in fermionic cold atoms on an optical lattice: A design
with a density functional formalism"}

\author{Sota Kitamura}
\affiliation{Department of Physics, University of Tokyo, Hongo, Tokyo 113-0033,
Japan}

\author{Naoto Tsuji}
\affiliation{Department of Physics, University of Tokyo, Hongo, Tokyo 113-0033,
Japan}

\author{Hideo Aoki}
\affiliation{Department of Physics, University of Tokyo, Hongo, Tokyo 113-0033,
Japan}

\maketitle
In this Supplemental Material, we 
first explain in Section A how to engineer the optical lattice potential Eqs.(2-4), which is essential for the realization of the present proposal.
Then, in order to clarify the basic idea, Section B describes how the tight-binding limit of the present system can be related to the checkerboard lattice model with a unitary transformation, Section C explains how the matrix elements for the hopping and interaction in the tight-binding model are estimated, and Section D defines the order parameter for the present system. 
In Section E, we elaborate the non-collinear formalism of the LSDA for cold atom systems from both theoretical and numerical aspects.
Section F shows detailed results of the DFT calculation obtained by changing the anisotropy parameter, and explains why the TMI phase is stabilized in particular cases.
Section G is devoted to the discussion and the proposal on the experimental detection of the topological signatures, where we show the simulated spectra of the light Bragg scattering for the present proposal, which highlights the topological edge modes.

\appendix
\section{Implementation of the lattice potential}

Here we indicate how we can realize the lattice potential given in
Eqs.~(2-4). We consider a situation where the laser electric fields
$\sum_{\omega}\bm{E}(\omega,\bm{r})e^{i\omega t}+\text{c.c.}$ are
imposed to atoms with a hyperfine structure. If all the laser frequencies
($\omega$'s) are off-resonant with the hyperfine splitting, the effect
of the laser field is represented as additional terms in the Hamiltonian
\cite{Deutsch1998-S,Dudarev2004-S}, a potential $\propto|\bm{E}(\omega,\bm{r})|^{2}$
along with a Zeeman term $\propto-i\bm{E}^{\ast}(\omega,\bm{r})\times\bm{E}(\omega,\bm{r})\cdot\hat{\bm{F}}$
with $\omega$-dependent coefficients, where $\hat{\bm{F}}$ is the
total angular momentum operator. For the present system we focus on
$F=1/2$ multiplets (then $\hat{\bm{F}}=\bm{s}_{\sigma\sigma^{\prime}}$).

\begin{figure}
\begin{centering}
\includegraphics[width=0.75\columnwidth]{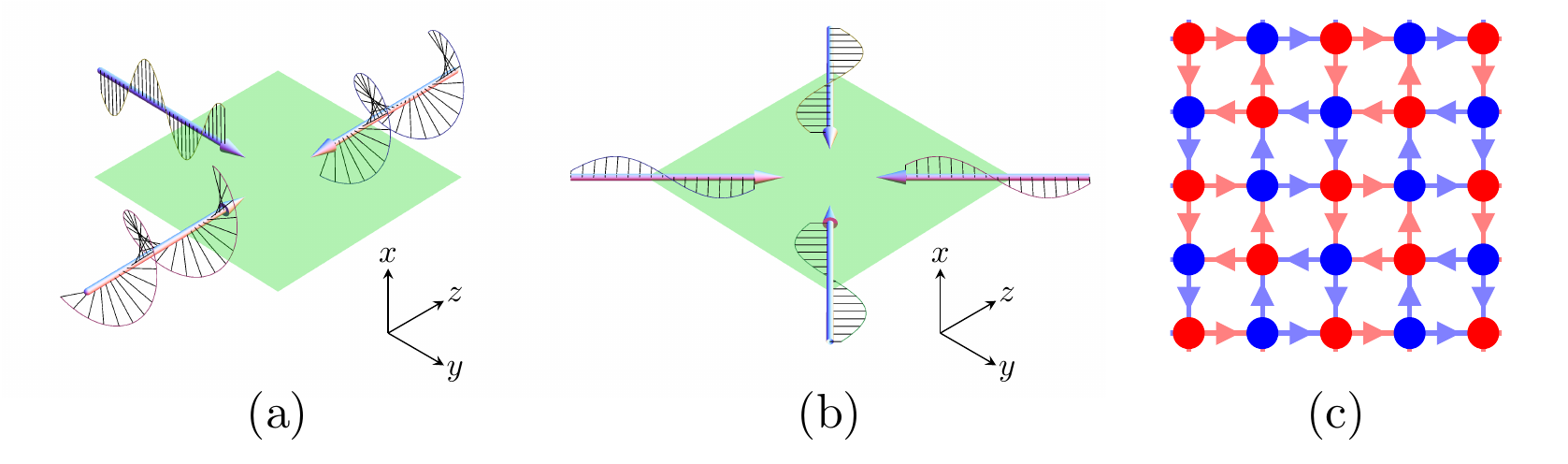}
\par\end{centering}

\protect\caption{\label{fig:flux} (a, b) Schematic pictures of the linearly and circularly
polarized laser configurations expressed by Eqs.~(\ref{eq:laserfield1},~\ref{eq:laserfield2}).
(c) Spatial pattern of the hopping introduced
by the mean-field decoupling, Eq.~(\ref{eq:fockterm}).  
Hopping along an arrow has an amplitude $iV\phi$, with
a hopping from $A$ to $B$ sites ($B$ to $A$ sites) depicted
in red (blue). }
\end{figure}

First, we consider a pair of confronting circularly-polarized lasers
along $z$-axis, and one linearly-polarized laser along $y$ 
(as schematically sketched in Fig.~S\ref{fig:flux}~(a)),
all with a frequency $\omega$, 
\begin{equation}
\bm{E}(\omega,\bm{r})\propto e^{\pi iy/d}\bm{e}_{x}+\sum_{\xi=\pm}\xi e^{\xi\pi iz/d}\dfrac{\bm{e}_{x}+i\xi\bm{e}_{y}}{\sqrt{2}},\label{eq:laserfield1}
\end{equation}
which can be shown to realize the $W$- and $B^{z}$-components of
$V_{\sigma\sigma^{\prime}}(\bm{r})$: 
\begin{gather}
|\bm{E}(\omega,\bm{r})|^{2}\propto3+\sqrt{2}\cos\dfrac{\pi(y-z)}{d}-\sqrt{2}\cos\dfrac{\pi(y+z)}{d},\, -i\bm{E}^{\ast}(\omega,\bm{r})\times\bm{E}(\omega,\bm{r})\propto\Bigl[\cos\dfrac{\pi(y-z)}{d}+\cos\dfrac{\pi(y+z)}{d}\Bigr]\bm{e}_{z}.
\end{gather}
We further superpose four linearly-polarized lasers 
(Fig.~S\ref{fig:flux}~(b)) of a frequency
$\omega^{\prime}=\omega/\sqrt{2}$ with a spatial part %
\footnote{In place of $\omega/\sqrt{2}$ we can employ $\omega\sqrt{2+p^{2}}/2$,
for which the laser field is obtained by substituting $y\rightarrow y+px$
in Eq.~(\ref{eq:laserfield2}). It has the desired form on the system
plane, $x=0$.%
}, 
\begin{equation}
\bm{E}(\omega^{\prime},\bm{r})\propto\sum_{\xi=\pm}\Bigl[e^{\xi\pi i(y-z)/2d}(\bm{e}_{y}+\bm{e}_{z})-\xi e^{\xi\pi i(y+z)/2d}(\bm{e}_{y}-\bm{e}_{z})\Bigr],\label{eq:laserfield2}
\end{equation}
which realizes the $W$- and $B^{x}$-components: 
\begin{gather}
|\bm{E}(\omega^{\prime},\bm{r})|^{2}\propto2+\cos\dfrac{\pi(y-z)}{d}-\cos\dfrac{\pi(y+z)}{d},\,
-i\bm{E}^{\ast}(\omega^{\prime},\bm{r})\times\bm{E}(\omega^{\prime},\bm{r})\propto\Bigl(\sin\dfrac{\pi y}{d}+\sin\dfrac{\pi z}{d}\Bigr)\bm{e}_{x}.
\end{gather}
If we combine laser fields in Eqs.~(\ref{eq:laserfield1},~\ref{eq:laserfield2}),
we end up with the desired Hamiltonian $\hat{H}_{\text{OL}}$, 
Eqs.(1-4) in the main text.

While we can change two strengths of the fields Eq.~(\ref{eq:laserfield1})
and (\ref{eq:laserfield2}), we have three parameters, $V_{A}$, $V_{B}$,
and $V_{C}$, to adjust in Eqs.~(2-4). 
If a tuning of the two strengths does not
attain the desired parameters, we can introduce additional, linearly-polarized
lasers, e.g., 
\begin{equation}
\bm{E}(\omega^{\prime\prime},\bm{r})\propto\sum_{\xi=\pm}\Bigl[e^{\xi\pi i(x/2+y+z)/2d}+\xi e^{\xi\pi i(x/2-y+z)/2d}\Bigr](\bm{e}_{z}-2\bm{e}_{x}),
\end{equation}
\begin{gather}
|\bm{E}(\omega^{\prime\prime},\bm{r})|^{2}\propto2-\cos\dfrac{\pi(y-z-x/2)}{d}+\cos\dfrac{\pi(y+z+x/2)}{d},\,
-i\bm{E}^{\ast}(\omega^{\prime\prime},\bm{r})\times\bm{E}(\omega^{\prime\prime},\bm{r})=0.
\end{gather}


\section{Equivalence of the tight-binding limit to the checkerboard lattice
model}

Here we describe how the present Hamiltonian $\hat{H}_{\text{OL}}$
is related to the checkerboard lattice model \cite{Sun2009-S} in
the tight-binding limit. As we have discussed in the main text, the
tight-binding limit is depicted in Fig.~1~(d), whose Hamiltonian reads 
\begin{align}
\hat{H}= & -\sum_{(i,j)\in A}(t_{A}\hat{a}_{i+1,j-1}^{\dagger}+t_{B}\hat{a}_{i+1,j+1}^{\dagger})\hat{a}_{i,j}+\text{h.c.}-\sum_{(i,j)\in B}(t_{B}\hat{b}_{i+1,j-1}^{\dagger}+t_{A}\hat{b}_{i+1,j+1}^{\dagger})\hat{b}_{i,j}+\text{h.c.}\nonumber \\
 & -t\sum_{(i,j)\in A}\sum_{\xi=\pm1}\xi(-1)^{j}(\hat{b}_{i+\xi,j}^{\dagger}+\hat{b}_{i,j+\xi}^{\dagger})\hat{a}_{i,j}+\text{h.c.}+V_1\sum_{(i,j)\in A}\sum_{(i^{\prime},j^{\prime})\in B}^{\text{n.n.}}\hat{a}_{i,j}^{\dagger}\hat{b}_{i^{\prime},j^{\prime}}^{\dagger}\hat{b}_{i^{\prime},j^{\prime}}\hat{a}_{i,j},\label{eq:latticemodel}
\end{align}
where $\hat{a}_{i,j}$ annihilates a spin-up ($A$ site) fermion at
$(y,z)=(i,j)$ with the lattice constant taken to be unity, 
while $\hat{b}_{i,j}$ is for a spin-down
($B$) fermion.

In order to relate Eq.~(\ref{eq:latticemodel}) to the checkerboard
lattice, we can perform a unitary transformation, 
\begin{gather}
\hat{a}_{i,j}^{\prime}=(-1)^{(i-j)/2}\hat{a}_{i,j},\,
\hat{b}_{i,j}^{\prime}=(-1)^{(i+j-1)/2}\hat{b}_{i,j}.\label{eq:unitary}
\end{gather}
Then the transformed model reads 
\begin{align}
\hat{H}= & -\sum_{(i,j)\in A}(-t_{A}\hat{a}_{i+1,j-1}^{\prime\dagger}+t_{B}\hat{a}_{i+1,j+1}^{\prime\dagger})\hat{a}_{i,j}^{\prime}+\text{h.c.}-\sum_{(i,j)\in B}(t_{B}\hat{b}_{i+1,j-1}^{\prime\dagger}-t_{A}\hat{b}_{i+1,j+1}^{\prime\dagger})\hat{b}_{i,j}^{\prime}+\text{h.c.}\nonumber \\
 & -t\sum_{(i,j)\in A}\sum_{\xi=\pm1}(\hat{b}_{i+\xi,j}^{\prime\dagger}+\hat{b}_{i,j+\xi}^{\prime\dagger})\hat{a}_{i,j}^{\prime}+\text{h.c.}+V_1\sum_{(i,j)\in A}\sum_{(i^{\prime},j^{\prime})\in B}^{\text{n.n.}}\hat{a}_{i,j}^{\prime\dagger}\hat{b}_{i^{\prime},j^{\prime}}^{\prime\dagger}\hat{b}_{i^{\prime},j^{\prime}}^{\prime}\hat{a}_{i,j}^{\prime},
\end{align}
which precisely coincides with the spinless checkerboard lattice
model, Fig.~1~(e) in the main text. In the derivation we have used
that $(-1)^{i-j}=1$ for $(i,j)\in A$.

\section{Qualitative estimation of model parameters in the tight-binding limit}

We can estimate effective model parameters for the proposed
system in the tight-binding limit, Eq.~(\ref{eq:latticemodel}), 
by approximating the Wannier orbitals by Gaussian functions and 
neglecting screening effects. We first 
consider the non-interacting case with 
$V_{C}=0$, for which the Wannier basis in this spin-diagonal
problem comprises 
\begin{gather}
\hat{\psi}_{\uparrow}(\bm{r})=\sum_{(i,j)\in A}w_{A}(\bm{r}-id\bm{e}_{y}-jd\bm{e}_{z})\hat{a}_{i,j},\,
\hat{\psi}_{\downarrow}(\bm{r})=\sum_{(i,j)\in B}w_{B}(\bm{r}-id\bm{e}_{y}-jd\bm{e}_{z})\hat{b}_{i,j}.
\end{gather}
If the
intensities of the lasers, $V_{A}$, $V_{B}$, and $V_{x}$, are sufficiently
strong, we can approximate the Wannier orbitals by Gaussian functions,
because the system can be approximated by harmonic oscillators.
Namely, 
\begin{gather}
w_{A}(\bm{r})\sim\Bigl(\dfrac{\pi^{6}V_{x}V_{A}V_{B}}{32E_{R}^{3}d^{12}}\Bigr)^{1/8}\exp\Bigl[-\dfrac{\sqrt{2}\pi^{2}}{8d^{2}}\Bigl(\sqrt{\dfrac{V_{x}}{E_{R}}}x^{2}+\sqrt{\dfrac{V_{A}}{E_{R}}}(y-z)^{2}+\sqrt{\dfrac{V_{B}}{E_{R}}}(y+z)^{2}\Bigr)\Bigr],\\
w_{B}(\bm{r})\sim\Bigl(\dfrac{\pi^{6}V_{x}V_{A}V_{B}}{32E_{R}^{3}d^{12}}\Bigr)^{1/8}\exp\Bigl[-\dfrac{\sqrt{2}\pi^{2}}{8d^{2}}\Bigl(\sqrt{\dfrac{V_{x}}{E_{R}}}x^{2}+\sqrt{\dfrac{V_{B}}{E_{R}}}(y-z)^{2}+\sqrt{\dfrac{V_{A}}{E_{R}}}(y+z)^{2}\Bigr)\Bigr].
\end{gather}
 We can then calculate $t$ and $V_1$ as 
\begin{align}
t & =\int d\bm{r}w_{A}^{\ast}(\bm{r})V_{C}\Bigl(\sin\dfrac{\pi y}{d}+\sin\dfrac{\pi z}{d}\Bigr)w_{B}(\bm{r}-d\bm{e}_{y})\nonumber \\
 & =2V_{C}\Bigl(\dfrac{V_{A}V_{B}}{V_{0}^{2}}\Bigr)^{1/4}\Bigl[1-\sin\dfrac{\pi(V_{B}-V_{A})}{2V_{0}}\Bigr]\exp\Bigl(-\sqrt{\dfrac{E_{R}}{2V_{0}}}-\dfrac{\pi^{2}}{2}\sqrt{\dfrac{V_{A}V_{B}}{2V_{0}E_{R}}}\Bigr),\\
V_1 & =\dfrac{4\pi\hbar^{2}}{M}a_{s}\int d\bm{r}|w_{A}(\bm{r})|^{2}|w_{B}(\bm{r}-d\bm{e}_{y})|^{2}\nonumber \\
 & =4E_{R}\dfrac{a_{s}}{d}\Bigl(2\pi^{2}\dfrac{V_{x}}{E_{R}}\Bigr)^{1/4}\sqrt{\dfrac{V_{A}V_{B}}{V_{0}E_{R}}}\exp\Bigl(-\pi^{2}\sqrt{\dfrac{V_{A}V_{B}}{2V_{0}E_{R}}}\Bigr),
\end{align}
where $V_{0}=(\sqrt{V_{A}}+\sqrt{V_{B}})^{2}$, while $t_{A}$ and
$t_{B}$ are estimated from the one-dimensional Mathieu equation as
\begin{align}
t_{A}  =\dfrac{4E_{R}}{\sqrt{\pi}}\Bigl(\dfrac{V_{A}}{2E_{R}}\Bigr)^{3/4}\exp\Bigl(-\sqrt{\dfrac{2V_{A}}{E_{R}}}\Bigr),\,
t_{B}  =\dfrac{4E_{R}}{\sqrt{\pi}}\Bigl(\dfrac{V_{B}}{2E_{R}}\Bigr)^{3/4}\exp\Bigl(-\sqrt{\dfrac{2V_{B}}{E_{R}}}\Bigr).
\end{align}

For instance, for the lattice parameters employed in the main text,
$V_{A}=0.8E_{R}$, $V_{B}=1.2E_{R}$, $V_{C}=0.25E_{R}$, and $a_{s}=0.25d$,
these estimations give $t=0.026E_{R}$, $t_{A}=0.320E_{R}$, $t_{B}=0.326E_{R}$,
and $V_1=0.059E_{R}$, although in this situation the approximation
will somewhat underestimate $t$ and $V_1$. However, 
we can expect that the proposed system is
weakly-correlated ($V_1/t_{A}<1$), for which the LSDA should provide accurate results.

\section{Order parameter in the mean-field description}

Here we discuss the order parameter of the optical lattice system
from the corresponding mean-field description of the tight-binding
checkerboard lattice. The order parameter of the Chern insulating
phase in the checkerboard lattice~\cite{Sun2009-S} is given as $\phi$,
where

\begin{equation}
\langle\hat{a}_{i,j}^{\prime\dagger}\hat{b}_{i^{\prime},j^{\prime}}^{\prime}\rangle=\begin{cases}
i\phi & {\rm for}\; (i^{\prime},j^{\prime})=(i\pm1,j),\\
-i\phi & {\rm for}\; (i^{\prime},j^{\prime})=(i,j\pm1).
\end{cases}
\end{equation}
We can readily go back to the tight-binding description of the optical
lattice with Eqs.~(\ref{eq:unitary}), and the
order parameter emerges as

\begin{equation}
\langle\hat{a}_{i,j}^{\dagger}\hat{b}_{i^{\prime},j^{\prime}}\rangle=\begin{cases}
i\phi(-1)^{j} & {\rm for}\; (i^{\prime},j^{\prime})=(i+1,j),(i,j-1),\\
-i\phi(-1)^{j} & {\rm for}\; (i^{\prime},j^{\prime})=(i,j+1),(i-1,j).
\end{cases}\label{eq:orderparameter}
\end{equation}

While this is the order parameter for the optical lattice system expressed
in terms of the tight-binding picture, we can introduce an alternative,
basis-independent observable appropriate to continuous problems. We
can start with an observation that the Fock term corresponding to
Eq.~(\ref{eq:orderparameter}) in the mean-field decoupling is 
\begin{equation}
-iV_1\phi\sum_{(i,j)\in A}\sum_{\xi=\pm1}\xi(-1)^{j}(\hat{b}_{i+\xi,j}^{\dagger}-\hat{b}_{i,j+\xi}^{\dagger})\hat{a}_{i,j}+\text{h.c.},\label{eq:fockterm}
\end{equation}
as depicted in Fig.~S\ref{fig:flux}~(c). As explained in the main
text, NN hopping with real amplitudes is obtained by the $B^{x}$-component,
while that with imaginary amplitudes is given by the $B^{y}$-component,
since it realizes spin-offdiagonal potential $V_{\uparrow\downarrow}=-V_{\downarrow\uparrow}=iB^{y}$.
The staggered pattern in Eq.~(\ref{eq:fockterm}) can be realized
by a sinusoidal 
\begin{gather}
B^{y}(\bm{r})\propto\sin\dfrac{\pi y}{d}-\sin\dfrac{\pi z}{d},\label{eq:staggered}
\end{gather}
which enables us to define the order parameter as the staggered magnetization
along $y$-axis, proportional to Eq.~(\ref{eq:staggered}): 
\begin{equation}
\sum_{\sigma,\sigma^{\prime}}\int d\bm{r}\hat{\psi}_{\sigma}^{\dagger}(\bm{r})\Bigl(\sin\dfrac{\pi z}{d}-\sin\dfrac{\pi y}{d}\Bigr)s_{\sigma\sigma^{\prime}}^{y}\hat{\psi}_{\sigma^{\prime}}(\bm{r}).
\end{equation}

\section{Non-collinear spin DFT for cold atom systems}

In this section we provide details of numerical calculations with the non-collinear spin DFT for cold atom systems.
Fundamental aspects of the DFT for cold atom systems are given in Ref.~\onlinecite{Ma2012} and its Supplementary Information for collinear cases, on which the following discussions are based.

We consider fermionic systems with a short-range interaction. While there are various types of short-range interactions, they can be characterized by the $s$-wave scattering length, $a_s$, and behave universally in the dilute limit. For instance, the hard-core interaction $U(\bm{r}-\bm{r}^\prime)=V\Theta(a_s-|\bm{r}-\bm{r}^\prime|)|_{V\rightarrow\infty}$ with $\Theta(x)$ being the step function, and the contact interaction $U(\bm{r}-\bm{r}^\prime)=(4\pi\hbar^2a_s/M)\delta(\bm{r}-\bm{r}^\prime)$ behave identically unless the density is too large. 

In the LSDA, we approximate the exchange-correlation energy at each position in real space by that of the homogeneous gas with corresponding spin densities. The homogeneous system can be simulated accurately with the diffusion Monte Carlo method, and the functional form obtained from the interpolation of simulation data is given in Supplementary Information of Ref.~\onlinecite{Ma2012}. Here the hard-core interaction is employed for the simulation.

In the collinear formalism of the spin DFT, with the polarization axis assumed to be collinear, arguments for the density functional are taken as the collinear spin densities, $n_\uparrow(\bm{r})$ and $n_\downarrow(\bm{r})$.  This contrasts with the non-collinear case, where the direction of the polarization is spatially-varying, and the arguments are the magnetization density vector $\bm{m}(\bm{r})$ along with the atomic density $n(\bm{r})$. In principle, the energy functional for non-collinear polarizations contains more information and have a complicated form. However, because the energy of the homogeneous gas is independent of the direction of the polarization, the LSDA functionals share their form between collinear and non-collinear cases: 
Namely, the non-collinear functional, $E^\text{(nc)}_{\text{HXC}}[n(\bm{r}),\bm{m}(\bm{r})]$, is represented by the collinear one, $E^\text{(c)}_{\text{HXC}}[n_{\uparrow}(\bm{r}),n_{\downarrow}(\bm{r})]$, as  $E^\text{(nc)}_{\text{HXC}}[n(\bm{r}),\bm{m}(\bm{r})] = E^\text{(c)}_{\text{HXC}}[(n(\bm{r})+|\bm{m}(\bm{r})|)/2,(n(\bm{r})-|\bm{m}(\bm{r})|)/2].$
Then the resulting Kohn-Sham potential reads 
\begin{equation}
V^\text{KS}_{\sigma\sigma^{\prime}}(\bm{r}:[n(\bm{r}),\bm{m}(\bm{r})])=V_{\sigma\sigma^{\prime}}(\bm{r})+\frac{{\delta}E^\text{(nc)}_{\text{HXC}}}{{\delta}n(\bm{r})}\delta_{\sigma\sigma^{\prime}}+\frac{{\delta}E^\text{(nc)}_{\text{HXC}}}{\delta|\bm{m}(\bm{r})|}\frac{1}{|\bm{m}(\bm{r})|}\bm{m}(\bm{r})\cdot\bm{s}_{\sigma\sigma^{\prime}}.
\end{equation}

In order to perform numerical calculations efficiently, we assume the ground state density and magnetization to be periodic with a period commensurate with that of the lattice.  Then we can adopt Bloch wavefunctions for each Kohn-Sham orbital, and we can then calculate the Kohn-Sham energy $\epsilon_{n\bm{k}}$ and orbital $\phi_{n\bm{k}\sigma}(\bm{r})$ for $n$-th band with crystal-momentum $\bm{k}$ by diagonalizing a one-body Kohn-Sham Hamiltonian,
\begin{equation}
[H^\text{KS}_{\bm{k}}(\bm{r})]_{\sigma\sigma^\prime}=-\frac{\hbar^2}{2M}\delta_{\sigma\sigma^\prime}(\bm{\nabla}+i\bm{k})^2+V^\text{KS}_{\sigma\sigma^{\prime}}(\bm{r}:[n(\bm{r}),\bm{m}(\bm{r})]). 
\end{equation}
The calculation is iterated until the atomic density 
$n(\bm{r})=\sum_{\sigma}\langle\hat{\psi}^\dagger_\sigma(\bm{r})\hat{\psi}_\sigma(\bm{r})\rangle=\sum_{n\bm{k}}\sum_{\sigma}f_{n\bm{k}}|\phi_{n\bm{k}\sigma}(\bm{r})|^2$ and the magnetization density  $\bm{m}(\bm{r})=\sum_{\sigma\sigma^\prime}\langle\hat{\psi}^\dagger_\sigma(\bm{r})\bm{s}_{\sigma\sigma^\prime}\hat{\psi}_{\sigma^\prime}(\bm{r})\rangle=\sum_{n\bm{k}}\sum_{\sigma\sigma^\prime}f_{n\bm{k}}\phi^\ast_{n\bm{k}\sigma}(\bm{r})\bm{s}_{\sigma\sigma^\prime}\phi_{n\bm{k}\sigma^\prime}(\bm{r})$ become self-consistent, where $f_{n\bm{k}}$ is the occupation.

Here we comment on the benefits of the method for the analysis of the present study.
The DFT formalism for many-body problems can achieve high accuracy despite its mean-field character (i.e. a one-body description in an effective medium), so that it is suitable for describing the TMI phase transition, which is originally proposed with a mean-field calculation. 
However, the TMI is realized mainly by the contribution from the Fock term. 
As a variational ansatz for the minimization of the total energy, the mean-field approximation decomposes the interaction energy  $(1/2){\int}d\bm{r}d\bm{r}^{\prime}\sum_{\sigma\sigma^\prime}U(\bm{r}-\bm{r}^{\prime})\langle\hat{\psi}_{\sigma}^{\dagger}(\bm{r})\hat{\psi}_{\sigma^\prime}^{\dagger}(\bm{r}^{\prime})\hat{\psi}_{\sigma^\prime}(\bm{r}^{\prime})\hat{\psi}_{\sigma}(\bm{r})\rangle$ into the Hartree and Fock terms as
\begin{gather}
E_\text{H}=\frac{1}{2}{\int}d\bm{r}d\bm{r}^{\prime}\sum_{\sigma\sigma^\prime} U(\bm{r}-\bm{r}^{\prime}) \langle\hat{\psi}_{\sigma}^{\dagger}(\bm{r})\hat{\psi}_{\sigma}(\bm{r})\rangle\langle{\psi}_{\sigma^\prime}^{\dagger}(\bm{r}^{\prime})\hat{\psi}_{\sigma^\prime}(\bm{r}^{\prime})\rangle=\frac{1}{2}{\int}d\bm{r}d\bm{r}^{\prime}\sum_{\sigma\sigma^\prime} U(\bm{r}-\bm{r}^{\prime}) n_{\sigma}(\bm{r})n_{\sigma^\prime}(\bm{r}^{\prime}),\\
E_\text{F}=-\frac{1}{2}{\int}d\bm{r}d\bm{r}^{\prime}\sum_{\sigma\sigma^\prime} U(\bm{r}-\bm{r}^{\prime}) \langle\hat{\psi}_{\sigma}^{\dagger}(\bm{r})\hat{\psi}_{\sigma^\prime}(\bm{r}^{\prime})\rangle\langle{\psi}_{\sigma^\prime}^{\dagger}(\bm{r}^{\prime})\hat{\psi}_{\sigma}(\bm{r})\rangle.
\end{gather}
While the Hartree term has an explicit spin-density-functional form, the Fock term is composed of non-local expectation values for $\bm{r}\neq\bm{r}^\prime$, and the form of the corresponding density-functional is unknown. For electronic systems, the Fock term in the DFT is thus not appropriately treated. By contrast, here we consider cold atom systems where the interaction is short-ranged: For the contact interaction $U(\bm{r}-\bm{r}^\prime)=(4\pi\hbar^2a_s/M)\delta(\bm{r}-\bm{r}^\prime)$, they are given as
\begin{gather}
E_\text{H}=\frac{4\pi\hbar^2}{M}a_s{\int}d\bm{r} n_{\uparrow}(\bm{r})n_{\downarrow}(\bm{r})=\frac{\pi\hbar^2}{M}a_s{\int}d\bm{r} \left[n^2(\bm{r})-m_z^2(\bm{r})\right],\\
E_\text{F}=-\frac{4\pi\hbar^2}{M}a_s{\int}d\bm{r}  |\langle\hat{\psi}_{\uparrow}^\dagger(\bm{r})\hat{\psi}_{\downarrow}(\bm{r})\rangle|^2=-\frac{\pi\hbar^2}{M}a_s{\int}d\bm{r} \left[m_x^2(\bm{r})+m_y^2(\bm{r})\right],
\end{gather}
where we can see that the both terms are explicitly represented as non-collinear density-functionals. Indeed, they are the first-order terms in the $a_s$ expansion of the LSDA functional. 

Let us also comment on the topological invariant. While in general the topological invariant for many-body systems is not trivially defined, here we define the topological invariant as that of the one-body Kohn-Sham Hamiltonian. This approach indeed works well in that the topological edge states emerge in the presence of boundaries unless the modulation of the effective potential, caused by that of the density near the boundary, significantly affects the bulk. Non-zero Hall conductivity is also realized, which should basically be characterized by the Chern number we adopt here.

\section{Enhancement of the order parameter by a lattice anisotropy}

The atomic density around $(y,z)=(\mp d/2,\pm d/2)$, which 
gives the upper limit for the order parameter as discussed in the main text, 
can be controlled 
by a lattice anisotropy as clearly depicted in Fig.~S\ref{fig:phase}, 
where $A-B$ is varied with $A+B$ and $C$ fixed to $2E_{R}$ and $0.25E_{R}$, respectively.
The phase boundary is here identified as the point at which the topological
gap opens with vanishing density of states, while the spontaneous
magnetization emerges prior to that to open a gap at the QBCP with
an initially overlapping bands for smaller repulsions.

\begin{center}
\begin{figure}
\begin{centering}
\includegraphics[width=0.6\columnwidth]{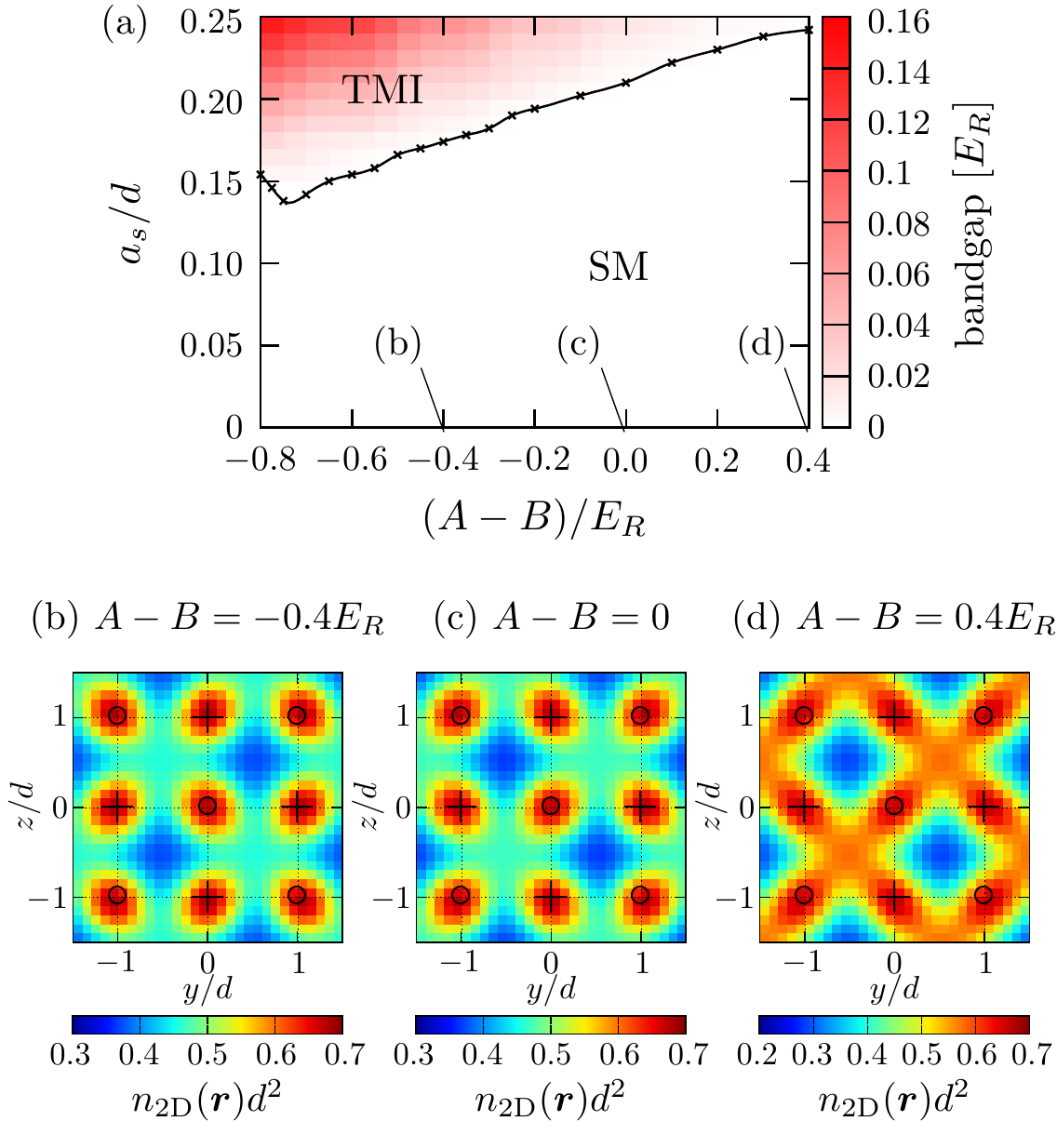}
\par\end{centering}

\protect\caption{\label{fig:phase} (a) A phase diagram against $A-B$ and $a_{s}$.
TMI stands for an insulating phase with
a spontaneous topological gap, while SM a semimetallic phase. 
The color represents the size of the
topological gap. (b-d) The atomic density of the non-interacting
system ($a_{s}=0$) for various values of $A-B$. }
\end{figure}

\par\end{center}

\section{Experimental detection}

The phase transition in the proposed
system is characterized by the staggered magnetization along $y$-axis 
as an order parameter. Hence the phase transition can be verified
by measuring the magnetization, or the accompanying energy gap. However,
in order to distinguish the phase transition from non-topological
phase transitions, the topological invariant, i.e. the Chern number
for the present system, should be measured in experiments. Although
this is challenging for cold-atom systems due to the lack of atomic
reservoirs required for a measurement of Hall currents in ordinary set-ups, various methods for identifying 
topological systems in cold-atoms are now being proposed.

Such proposals can be categorized into three: (1) Observation
of characteristic dynamics of topological systems, e.g., drift
of wave packets in 
the Bloch oscillation~\cite{Price2012,Jotzu2014,Dauphin2013},
or the propagation of the edge states along the boundary~\cite{Goldman2013};
(2) Direct evaluation of the winding number or the Chern density
from the spin-resolved time-of-flight images~\cite{Alba2013,Deng2014,Lisle2014,Hauke2014};
(3) Detection of the gapless excitations derived from edge states 
with spectroscopic methods~\cite{Stamper-Kurn1999,Liu2010,Goldman2012}. Here we discuss
the applicability of these methods to the present system.

(1) Topological Mott insulators are different from ordinary topological
systems in that the topological nature is interaction-driven, 
so that their topological character
is sensitive to filling of the system. Therefore, the observation of the 
wave-packet dynamics may be difficult for the present system: We have
to maintain the filling of system while making the wavefunction localized
at the same time.  

The proposal given in Ref.~\onlinecite{Dauphin2013}, where the half-filled
system is suddenly released from a trapping potential and the drift of 
its center-of-mass 
in a potential gradient reflects the Chern
number, satisfies this requirement and may be applicable to the present
system, although the signal will decay 
as the system spreads and the order decays. 
The topological gap should be large enough 
to prevent the potential gradient from mixing the bands 
and the hopping amplitudes from spreading of the system, 
which might be challenging.

(2)  The time-of-flight (ToF) measurement provides momentum-resolved (i.e.
eigenstate-resolved) information on the system. In particular, for the optical honeycomb lattice system with 
the sublattice degree of freedom implemented by the spin structure, 
the spin-resolved image gives a spin direction of each eigenstate
in momentum space, whose winding number coincides with the topological invariant
of the system~\cite{Alba2013}.

While the present system is constructed in a similar manner, 
the ToF image would not provide the Chern number.  
This is due to the difference in the structure of eigenstates:
In order to evaluate the Chern number of the present system, not only $\langle \hat{\psi}^\dagger_\sigma(\bm{p}) \hat{\psi}_\sigma(\bm{p}) \rangle$ but also $\langle \hat{\psi}^\dagger_\uparrow(\bm{p}) \hat{\psi}_\downarrow(\bm{p}+\bm{Q}) \rangle$ is required in the tight-binding limit, where $\bm{Q}=(0,\pi/d)$. 
Moreover, we employ the shallow lattice potential, where the eigenstates is more complicated, and the Berry curvature is derived from $\langle \hat{\psi}^\dagger_\sigma(\bm{p}) \hat{\psi}_{\sigma^\prime}(\bm{p}+\bm{P}) \rangle$ with various $\bm{P}$'s e.g. $(\pi/d,\pm \pi/d)$, so that the ToF images should be inadequate for detecting the Chern number properly.

(3) Now, a clearest way for identifying the TMI phase 
is to 
detect the gapless excitations due to the edge states, 
which are inherent in topological phases. The edge-state excitations can be
probed with various spectroscopic methods for cold-atom systems, and here we propose the light 
Bragg spectroscopy~\cite{Stamper-Kurn1999,Liu2010} 
should be most suitable and promising, and 
present a simulated spectrum.  
This way we should be able to optically detect the edge-modes, not only along the edge of the entire system, but
also around 
the boundaries of phase domains that may be present.

In the framework of the Bragg spectroscopy, we perturb the
system with a pair of probe lasers having wave-numbers 
$\bm{k}_1$ and $\bm{k}_2$, with the perturbative Hamiltonian
\begin{equation}
\hat{V}_{\text{Bragg}}(t)=\hbar\Omega\sum_{\sigma}\int d\bm{r}\hat{\psi}_{\sigma}^{\dagger}(\bm{r})\cos(qy-\omega t)\hat{\psi}_{\sigma}(\bm{r}).
\end{equation}
Here $\bm{q} = \bm{k}_2 - \bm{k}_1$ and $\omega$ are respectively the difference in the wave-number and
the frequency of the lasers, and we take the sample edges 
to be normal to the 
$z$ axis, while periodic along $y$. The light Bragg spectroscopy directly measures the
dynamical structure factor $S(q,\omega)$, which should reflect 
the edge-mode dispersions. 
In the present model calculation 
we impose a square well potential $\hat{V}_{\text{well}}$
with a width $80d$ along $z$.  For simplicity, we approximate
the total many-body system with the Kohn-Sham potential for
the bulk system, i.e., 
\begin{equation}
\hat{H}=\hat{H}_{\text{KS}}[n_{\text{bulk}}(\bm{r}),\bm{m}_{\text{bulk}}(\bm{r})]+\hat{V}_{\text{well}}+\hat{V}_{\text{Bragg}}(t).
\end{equation}

For the finite-width system we show the band structure in the non-interacting case 
(Fig.~S\ref{fig:bragg}~(a)) and the TMI phase (Fig.~S\ref{fig:bragg}~(b)).
In the TMI phase the edges modes appear within the gap, 
which comprise left- and right-edge states as we can identify 
from their color-coded center-of-mass coordinates 
\footnote{Even in the non-interacting case, the edge modes appear, but they sustain no chiral transports. %
}.  
The light Bragg spectrum is displayed 
in Fig.~S\ref{fig:bragg}~(c) 
 for the non-interacting case and in Fig.~S\ref{fig:bragg}~(d)  for the TMI phase. In the former the spectrum simply 
reflects the gapless semimetal. 
By a sharp contrast, the light Bragg spectrum for the 
TMI clearly exhibits edge-mode excitations that start from zero energy at $q=0$ up to the topological band 
edges. These come from intra-edge-mode excitations 
that delineate the edge modes within the gap. 
We can further confirm the gapless
excitation to be originating from edge states, by irradiating only
the one side of the sample. Figure ~S\ref{fig:bragg}~(e) shows the spectrum
in the TMI phase when only $z<0$ is illuminated, where 
a contribution
from the edge states at $z\sim40d$ (red curve in Fig.~S\ref{fig:bragg}~(b)) are seen to 
disappear, while the other mode is left intact.  
Thus, if we can make the potential well sharp enough, 
the topological edge modes are expected to be detectable in this manner.

\begin{center}
\begin{figure}
\begin{centering}
\includegraphics[width=0.8\columnwidth]{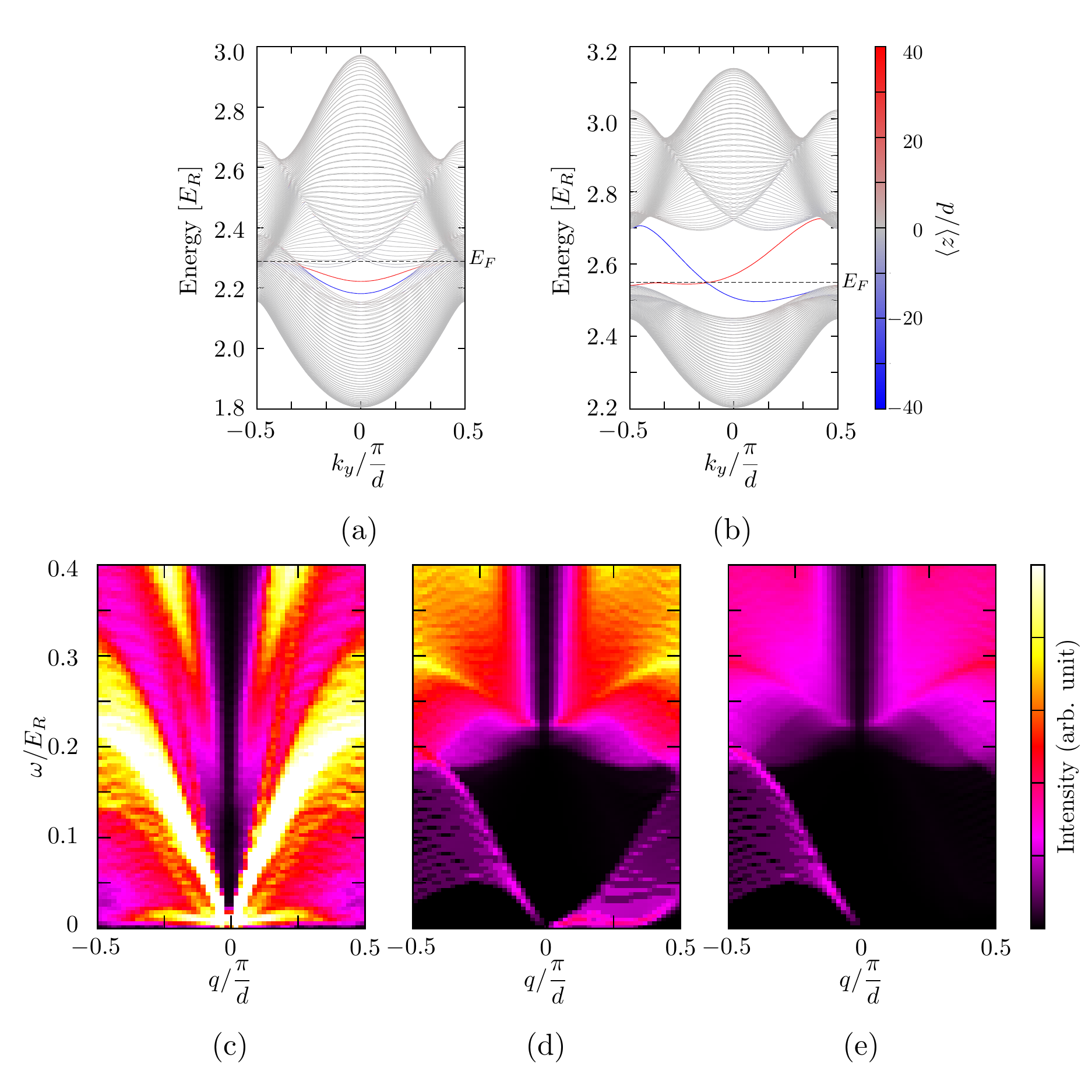} 
\par\end{centering}

\protect\caption{\label{fig:bragg} (a, b) Structure of the lowest four bands for the system with a finite width along 
$z$ in the non-interacting case (a)
and the interacting case with $a_{s}=0.25d$ (b). We set $(V_{A},V_{B},V_{C})=(0.6E_{R},1.4E_{R},0.25E_{R})$
in both cases. The dotted lines indicate the Fermi level $E_F$, while the
colored lines represent edge modes with a color-coded center-of-mass $z$ coordinate.  
(c-e) The simulated spectra of the light Bragg scattering for the
non-interacting (c) and interacting (d, e) cases. For (c, d) the
whole system is irradiated by the probe laser, while only the $z<0$
region is irradiated in (e).}
\end{figure}

\par\end{center}

\newcommand{\PRL}[3]{\href{http://link.aps.org/abstract/PRL/v#1/e#2}{Phys. Rev. Lett. \textbf{#1},#2 (#3)}}
\newcommand{\PRA}[3]{\href{http://link.aps.org/abstract/PRA/v#1/e#2}{Phys. Rev. A \textbf{#1}, #2 (#3)}}
\newcommand{\PRB}[3]{\href{http://link.aps.org/abstract/PRB/v#1/e#2}{Phys. Rev. B \textbf{#1}, #2 (#3)}}
\newcommand{\JPSJ}[3]{\href{http://dx.doi.org/10.1143/JPSJ.#1.#2}{J. Phys. Soc. Jpn. \textbf{#1}, #2 (#3)}}
\newcommand{\arxiv}[1]{\href{http://arxiv.org/abs/#1}{arXiv:#1}}
\newcommand{\RMP}[3]{\href{http://link.aps.org/abstract/RMP/v#1/p#2}{Rev. Mod. Phys. \textbf{#1}, #2 (#3)}}